# Pollux: high-resolution precision spectroscopy and polarimetry for the Habitable Worlds Observatory

David Le Mignant*[a], Coralie Neiner[b], Jean-Claude Bouret[a], Luca Fossati[c], Eduard Muslimov[a,d], Kjetil Dohlen[a], Anne Bonnefoi[a], Jean-Michel Reess[b], Mélissa Alliguié[a], Sreejith A. G.[c] , Aquib Moin[e], Adrien Girardot[b], for the Pollux consortium.

[a]LAM (CNRS, CNES, Aix-Marseille Université), Marseille, France, [b]LIRA, Observatoire de Paris, CNRS, PSL Université, Sorbonne Université, Université Paris Cité, CY Cergy Paris Université, Meudon, France, [c]Space Research Institute, Austrian Academy of Sciences, Graz, Austria, [d]Department of Physics, University of Oxford, Oxford, UK, [e]United Arab Emirates University, Al Ain, United Arab Emirates

## ABSTRACT

Pollux is a high-resolution spectrograph and spectropolarimeter (R ~ 65,000 to 100,000) covering a spectral range from 100 nm to 1,750 nm, proposed by a European consortium to equip NASA's Habitable Worlds Observatory (HWO). This instrument aims to revolutionize the study of stellar and (exo)planetary systems, as well as cosmic ecosystems, by combining high spectral resolution, broad and simultaneous spectral coverage, temporal stability, and unique UV spectropolarimetric capabilities, thus opening a new parameter space for astrophysics.

## 1. INTRODUCTION

The Habitable Worlds Observatory (HWO) is a flagship NASA mission planned for launch in the early 2040s, designed as a multi-purpose observatory covering all domains of astrophysics, with a particular focus on detecting and characterizing habitable exoplanets. HWO follows in the footsteps of major space observatories such as Hubble and JWST, including an extended spectral coverage from UV to near infrared (NIR) and unprecedented instrumental capabilities (see Dressing [1] and Feinberg [2]).

Pollux is proposed by Neiner and Bouret [3] as one of HWO's key instruments, complementing other planned instruments such as a coronagraph, a high-resolution imager, a UV multi-object spectrograph (MOS), and an integral field unit (IFU). Its primary goal is to provide high-resolution spectroscopic and spectropolarimetric observations (R ~ 65,000 to 100,000) simultaneously across an optimized spectral range (100 nm – 1,750 nm), with temporal stability and sensitivity to perform transformational astrophysics.

This paper is structured as follows: Section 2 presents Pollux's science objectives, highlighting its capability to address major astrophysical questions, from star formation and exoplanet characterization to the study of cosmic ecosystems. Section 3 details the high-level requirements, including the instrument's functionalities, operation modes, and performance specifications. Section 4 describes the optical architecture, focusing on the distinct light paths for the FUV and MUV-to-NIR channels, as well as the design of the spectrographs and polarimeters. Section 5 discusses the performance of Pollux, including the signal-to-noise calculator and the instrument's expected sensitivity. Section 6 outlines the consortium and collaborations driving the project, while Section 7 covers the ongoing concept maturation activities, including technological advancements in detectors, coatings, polarimeters, and gratings. Finally, Section 8 presents future work and the next steps for Pollux, including funding efforts and international partnerships.

* david.lemignant@lam.fr ; hwo-pollux.eu

## 2. POLLUX SCIENCE OBJECTIVES

Pollux aims to address major astrophysical questions, including star formation, evolution, and chemical enrichment. Pollux will investigate the processes of star formation and the universe's chemical history, including searching for the first metal-free stars, understanding how massive stars create heavy elements and influence their surroundings through mass loss and rotation, and tracing how those elements enrich subsequent stellar generations. It will also study low-mass stars like white dwarfs to probe the composition of destroyed exoplanetary material.

A critical objective is to understand how planetary systems form and evolve. Pollux will study young planetary systems, debris disks, and the role of magnetic fields in shaping these environments to understand how planets get their initial compositions and orbits.

The instrument will perform detailed characterization of exoplanet atmospheres to determine their composition, dynamics, and loss. A key goal is to conduct the first definitive searches for and mapping of exoplanetary magnetic fields by detecting auroral emissions, which are crucial for atmospheric shielding and long-term habitability.

Pollux will study objects within our Solar System to understand their physical properties and the origins of life's building blocks. It will analyze asteroids and comets to determine their surface composition and perform spectropolarimetry of icy ocean worlds like Europa, Ganymede, and Enceladus to diagnose surface properties, identify regions of plume activity, and study their interaction with magnetospheres.

Pollux aims to explore the interstellar medium and the broader ecosystem of galaxies. It will map the 3D structure of magnetic fields in galactic environments, trace interstellar dust properties, and provide a comprehensive census of the different gas phases within and around galaxies.

The instrument will test the fundamental laws of physics and probe the early universe. By observing distant quasars and white dwarfs, it will test for variations in physical constants and measure primordial element abundances, providing new constraints for cosmological models.

Pollux will investigate the nature of the most energetic cosmic events, such as supermassive black hole binaries and nuclear transients like tidal disruption events. Its unique spectropolarimetric capabilities will help map the structure and physics of these phenomena.

Its most innovative features are its long, simultaneous wavelength coverage and its optional spectropolarimetric capability, which for the first time will enable in-depth studies of transient events and provide access to the study of magnetic fields and scattering processes.

Pollux's science cases are organized into three main themes—stars, (exo)planets, and cosmic ecosystems—aligned with HWO's priorities [3]. The Pollux project office issued a call for science cases to better delineate Pollux science and clearly identify science requirements. The call is still open, and anyone is invited to submit a science case through a web form (https://forms.gle/uvRg8Md2wwxDCwf26). The science themes cover a wide range of topics, from physical processes in our solar system to magnetic stars, exoplanet characterization, and cosmic ecosystem exploration.

### Solar system.

Pollux will extend HWO science into our own Solar System through spectropolarimetric studies of asteroids, comets, and icy ocean worlds. For asteroids and comets, it will use wavelength- and phase-dependent linear polarization to characterize surface properties such as grain size, composition, and albedo, as well as particulate cloud properties; in comets, it may identify amino acids through enantiomer-specific UV polarization signatures, offering constraints on prebiotic asymmetry. For ocean worlds such as Europa, Ganymede, and Enceladus, Pollux will obtain spatially resolved spectropolarimetry of reflected sunlight to diagnose ice grain size, porosity, texture, and contamination by salts or organics, helping identify regions linked to plume activity or subsurface exchange. It will also use UV spectropolarimetry of oxygen lines in these moons to probe magnetospheric interactions and precipitating electron populations, while also constraining haze particle properties on Titan.

### Stars.

Pollux will address stellar evolution from the first generations of stars to evolved remnants and massive stars. It will search for surviving low-mass, metal-free stars and derive precise abundances of elements such as carbon, magnesium, aluminum, silicon, and iron in later generations of cool FGK stars, thereby tracing early chemical enrichment and the yields of the first stellar populations. Its UV access will be especially important for r-process diagnostics, because many heavy-element absorption lines lie predominantly in the UV, increasing the detectable heavy elements in cool stars compared with optical-only observations. Pollux will also study massive O, B, and Wolf-Rayet stars across different metallicities, using high-resolution UV spectroscopy and spectropolarimetry to constrain stellar parameters, mass loss, clumping, rotation, wind asymmetries, and binary-shaped structures. In addition, it will use polluted white dwarfs to infer the bulk composition of extrasolar planetary material and to test possible variations in fundamental constants in strong gravitational fields.

### Exoplanets.

Pollux will be instrumental in directly characterizing exoplanet atmospheres, including atmospheric escape, mass loss, composition, dynamics, aerosols, and magnetic environments. By simultaneously observing diagnostics from the UV, optical, and near-infrared, such as Ly-alpha, Mg II, Fe I, Fe II, H-alpha, and the He I metastable triplet, it will provide a multi-layered view of planetary upper atmospheres and clarify how stellar high-energy irradiation drives atmospheric loss. For sub-Neptunes, high-resolution spectroscopy will determine molecular abundances such as water, methane, and ammonia, as well as bulk atmospheric metallicity, helping distinguish water-dominated worlds, hydrogen-rich mini-Neptunes, and potentially habitable Hycean worlds. Pollux will also map atmospheric velocity fields and wind speeds in giant exoplanets, constrain aerosol scattering and absorption across UV to NIR wavelengths, and search for exoplanetary magnetospheres through auroral UV emissions and star-planet magnetic interaction signatures.

### Cosmic ecosystems.

Pollux will provide a comprehensive view of the interstellar medium and galactic environments by accessing atomic and molecular transitions from the far-UV to the near-infrared, including species such as CO, $H_2$, C I, C II, O I, Mg II, Fe II, Si II, C IV, Si IV, and O VI. This will enable detailed studies of gas phases in and around galaxies, element depletion onto dust grains, molecular gas formation, ionization, shocks, and the behavior of the ISM as a dynamic, unstable, magnetized medium. Its spectropolarimetric capabilities will support three-dimensional interstellar magnetic-field tomography and dust-grain physics through broadband Stokes measurements, helping reconstruct plane-of-sky magnetic fields and diagnose dust-grain alignment and size distributions. Pollux will also connect Milky Way UV-bright stellar populations to the interpretation of UV upturn galaxies and broader galaxy evolution, building more complete stellar population libraries and helping reconstruct star-formation histories out to intermediate redshifts.

## 3. HIGH-LEVEL REQUIREMENTS

Pollux' science cases require a high-resolution spectrograph with optional polarimetric capability from the FUV to the NIR. The functionalities, operation modes, and performance requirements have been discussed within the science working groups since early 2025 and within the consortium during various meetings, including the Pollux consortium meetings in Paris (2025) and Leuven (2026). The Pollux optical design by Muslimov [4,5] has evolved accordingly.

| | | simultaneous | | | |
|---|---|---|---|---|---|
| | FUV | MUV | NUV | OPT | NIR |
| Wavelength Range (nm) | 100-123 | 120-236 | 236-438 | 438-875 | 875-1750 |
| Spectral Resolution | 100k | 100k | 100k | 65k | 65k |
| Point source spectropolarimetry | Yes | Yes | Yes | Yes | Yes |
| Point source spectroscopy | No | Yes | Yes | Yes | Yes |
| Slit spectroscopy (≤3 arcsec) | No | Yes | Yes | No | No |

*Figure 1: Pollux proposed functionalities and operation modes.*

Pollux covers a large wavelength range from 100 nm to 1750 nm across five distinct channels: FUV, MUV, NUV, OPT, and NIR. Each channel features its own echelle spectrograph and retractable polarimeter. The FUV channel is separate from the other four channels (MUV, NUV, OPT, and NIR), which can be operated simultaneously. The required spectral resolution, derived from the science cases, varies from 100,000 in the UV channels (FUV, MUV, NUV) to 65,000 in the OPT and NIR channels. The retractable polarimeters in each channel measure the full Stokes parameters (I, Q, U, V), providing both circular and linear polarization data with a precision of 10^-4, with an ambitious goal of 10^-6 for specific science cases, such as the detection of weak magnetic fields in exoplanetary atmospheres. This precision has been technically validated through laboratory tests of the UV and FUV polarimeters, as presented by Neiner and Girardot ([6,7,8]). The exact wavelength range and spectral resolution for each channel will continue to evolve as the spectrograph optical design matures, following the trade-space analyses.

Four primary operation modes are enabled by Pollux. The first mode is point-source high-resolution spectropolarimetry in the FUV channel (100–123 nm), which operates at a spectral resolution of R ≥ 100,000. This mode is designed to resolve narrow emission and absorption lines critical for studying stellar and interstellar medium properties. The FUV channel is equipped with a non-retractable dedicated polarimeter. However, the engineering team has identified an optical solution to enable a pure spectroscopy mode without moving the polarimeter (see Sect.5 of [5]), aiming to improve the signal-to-noise ratio (SNR) for spectroscopic observations that would suit several key science cases.

The second available mode is point-source spectroscopy, simultaneously covering the MUV, NUV, OPT, and NIR channels. This mode offers a spectral resolution of R ≥ 100,000 in the MUV and NUV channels and R ≥ 65,000 in the OPT and NIR channels. This configuration allows for comprehensive studies of a broad range of science cases. This mode also enables the user to choose the pin-hole size to further enhance flux stability over time.
The third mode is like the second but focuses on point-source spectropolarimetry with the same spectral resolution as mode 2. This mode is optimized for observations requiring polarimetric observations, for example to study magnetic fields and scattering processes.

The fourth and final mode enables slit spectroscopy in the MUV and NUV spectral windows at a spectral resolution of ≥ 100,000. Given the current design, the maximum available on-sky slit length could be of 4” in the MUV and 9” in the NUV [5]. This mode is particularly suited for solar system studies and cosmic ecosystem science cases, such as the investigation of aurorae on giant planets.

## 4. OPTICAL ARCHITECTURE

Pollux’s optical architecture begins at the HWO telescope’s M3/M4 relay. There are two very distinct light-paths: the FUV-only arm and the MUV-to-NIR arm, corresponding to two distinct locations (and telescope pointings) within the HWO Pollux field of view provided by the telescope.

Pollux optical design progress has been reported in several papers by Muslimov [4,5].

The FUV channel (100–123 nm) is designed for high throughput high-resolution spectropolarimetry at a spectral resolution of R ≥ 100,000. The FUV channel transmission budget is dominated by the polarimeter transmission, though the newly achieved spectroscopic option boosts the effective area significantly. Transmission materials are unavailable below 120 nm, so the polarimeter uses SiC mirrors [7,8].

Pollux optical design, optimized by Muslimov et al. [4,5], ensures minimal light loss and maximal efficiency across the 120–1750 nm spectral range. The beam passes through two separate dichroic elements, which split the light into the four spectral channels: MUV (120–236 nm), NUV (236–438 nm), OPT (438–875 nm), and NIR (875–1750 nm). The exact location of the wavelengths separating the four channels is still to be decided.

The MUV channel now starts at 120 nm instead of 101 nm, a change introduced in May 2026 to simplify the first dichroic and the MUV grating design. This adjustment adds the constrain that the 100–120 nm range is no longer simultaneous with the rest of the spectrum, even in spectroscopy mode. The spectral resolution is set to 100,000 for the FUV, MUV, and NUV channels, enabled by the reduced MUV range, while the OPT and NIR channels operate at 65,000, which is sufficient for the science goals. Each channel includes a collimator, echelle grating, and camera, with the gratings optimized for their respective spectral ranges as detailed in [4].

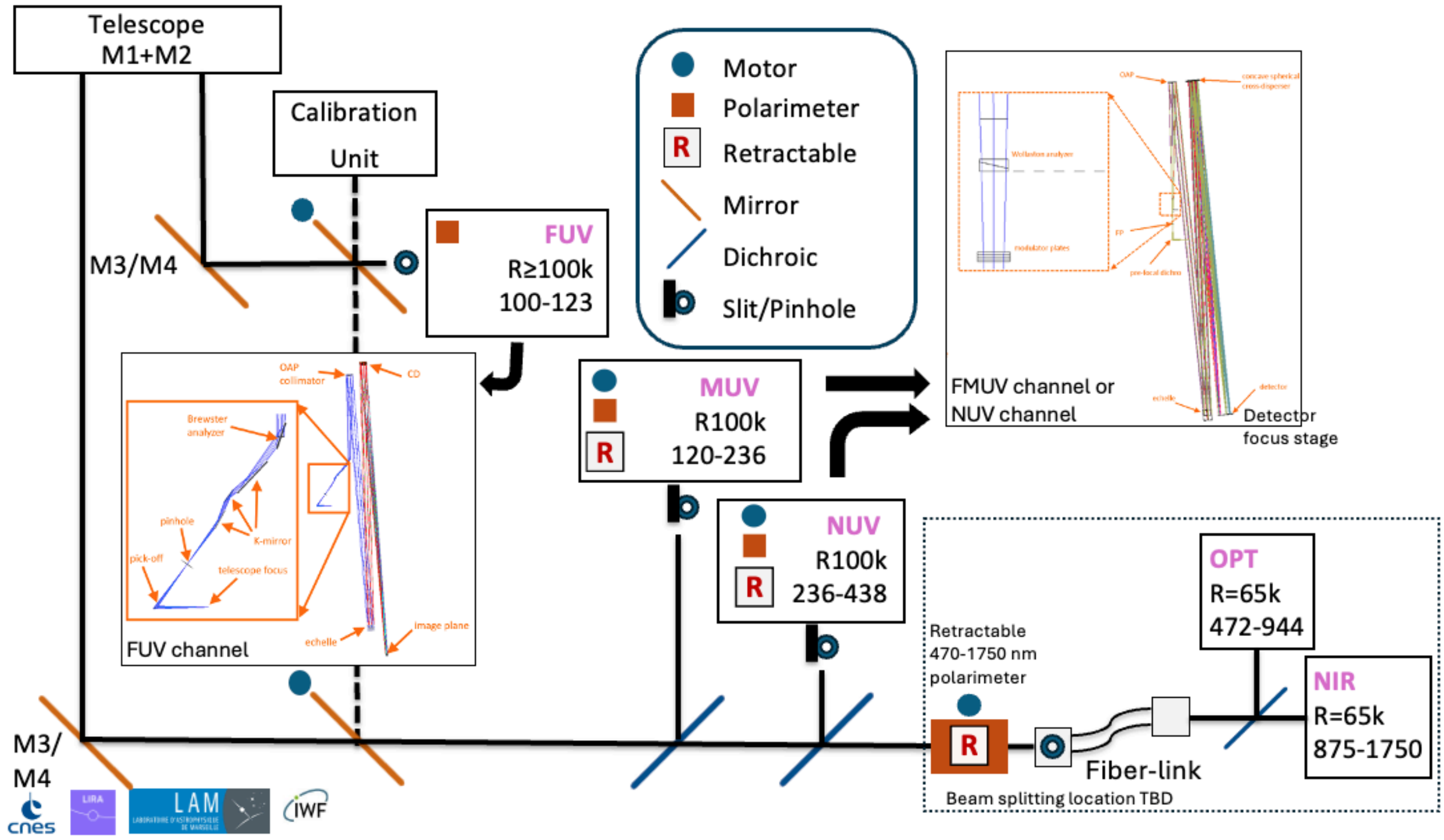

*Figure 2: Pollux optical layout. The exact sequence in the optical path of the retractable polarimeters and fiber link connections of the OPT and NIR is still to be discussed and what shown here is not representative of the final configuration.*

The detectors for each channel are tailored to their spectral ranges, with the FUV detector choice (CMOS or MCP) still under evaluation including an update on quantum efficiency measurements expected in 2026. The optical coatings (e.g., MgF2 for UV, Al + SiO2 for visible) are optimized to maximize transmission and minimize losses. The end-to-end transmission of the system remains a key focus through our studies.

The MUV and NUV channels retain an on-sky of maximum 4" and 9" arcsec respectively slit for some stellar, solar system and cosmic ecosystem science cases, allowing for slit spectroscopy. The optical and near-infrared channels are optimized for point-source spectroscopy and polarimetry, with no slit spectroscopy capability. These two channels are coupled via a fiber-injection module. The design for these channels has yet to be developed and will be based on existing spectrographs and spectropolarimeters with proven technologies. The use of a fiber link connection has been introduced to ease packaging of the instrument, but this choice will be re-evaluated once the size of the instrument box will become available.

## 5. PERFORMANCE

On the basis of the current optical design, the Pollux consortium developed a preliminary version of the signal-to-noise calculator. The application is being developed within the environment of the existing HWO-tools and extends it to the Pollux arms with a user-friendly graphical interface that enables users to specify observational parameters, execute calculations in real time, and visualize results through dynamic plots and tables. A key design goal of the software is adaptability to the evolving HWO telescope and Pollux instrumental architecture. As the observatory configuration continues to mature, the software is structured to accommodate modifications to the optical layout and instrument design with minimal effort. The framework is also designed to rapidly incorporate updated component performance data, including detector quantum efficiencies, reflectivity, and results from optical ray-tracing analyses, enabling timely reassessment of instrument capabilities throughout the development cycle. The software is implemented in Python and serves both as a research tool for assessing observational feasibility and as a planning tool for understanding the factors that govern measurement sensitivity to help and assist instrument design choices. Figure 3 shows Pollux effective areas from the FUV to NIR wavelengths.

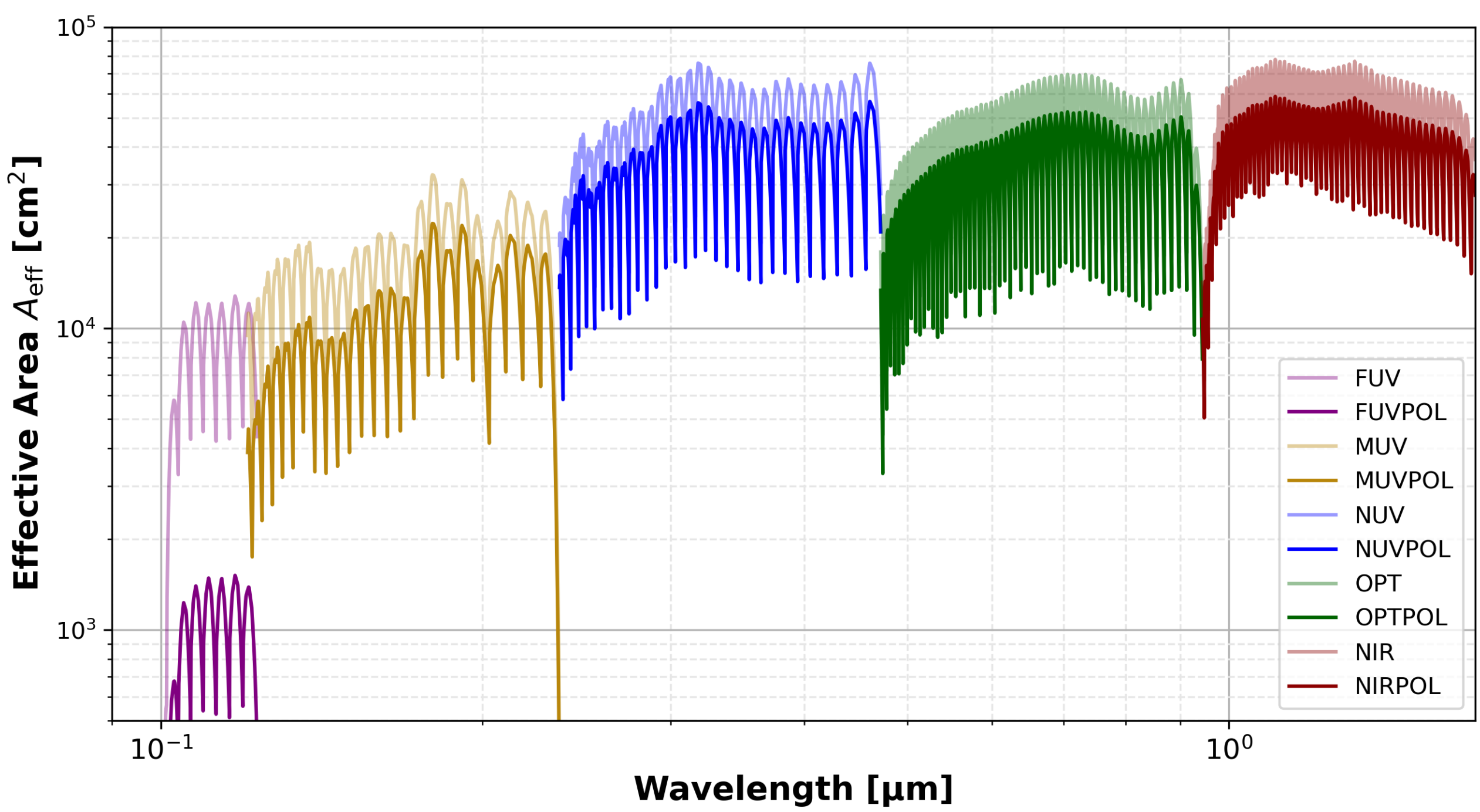

Figure 3. Pollux effective area curves.

## 6. CONSORTIUM

The three science working groups are led by their respective coordinator and the consortium has initiated a science mailing list that is now including 150 scientists. These European and international scientists are critical to explore and develop the science cases as well as initiate the dialog with the national space agencies.
In parallel, the Pollux project office is building up a European consortium to propose the Pollux instrument for HWO, and they look for expanding the consortium to other European countries. The Pollux project office also welcome any other international collaborations in the Americas, Asia, or Oceania. Possible contributions to the Pollux instrument payload have been primarily discussed with European countries, each bringing specialized expertise. Industrial partners have also been engaged to support the technological maturation of the instrument. Anyone interested to join the consortium is encouraged to contact the authors.

All these discussions are still in a preparatory phase, and no official contract nor memorandum of understanding has been signed yet. Pollux PIs are French and the French space agency is already supporting the project through the maturation and study phase.

## 7. CONCEPT MATURATION

Tuttle et al. [9] have presented the current state of a collection of UV technologies in preparation for the HWO. The concept maturation phase of Pollux focuses on advancing the instrument's technological readiness to align with HWO's timeline, targeting TRL 5 by the end of 2028. The current optical architecture and optical design provide a baseline for the functional and performance specifications of the various components. Key maturation activities have been identified in various areas: detectors, coatings, polarimeters, gratings, dichroics, M3/M4 optical relay and calibrations.

Pollux current design has 5 channels FUV, MUV, NUV, VIS and NIR. The trade-off study for the Pollux detectors is conducted by Jesper Skottfelt from the Center for Electronic Imaging, The Open University UK [10]. FUV and MUV range has two competing technologies detectors (MCP multichannel plates and CMOS image sensors – CIS). CMOS image sensors will be preferred for the NUV and VIS channels. Two competing technologies H4RGs and eAPDs (both based on Mercury Cadmium Telluride) may be proposed for the NIR channels. Many parameters are being considered for the trade-off evaluation including heritage & technology maturity, quantum efficiency optimization in the considered waveband, read noise and dark current, dynamic range, pixel size, array dimension (up to 9k x 9k) and modularity, run-time operation temperature for detectors and electronics & cooling requirements, power usage, operations and calibrations.

Solutions for UV polarimeters are in development at the Paris Observatory by Neiner 's group [6] and are currently implemented in Pollux optical design. Pollux's polarimeter development [7,8] focuses on achieving a precision of 10^-5, as required for its scientific objectives. The MUV-NUV polarimeter test bench, using a deuterium lamp (115–300 nm), has been successfully implemented, with initial results showing good agreement between theoretical Mueller calculations and experimental measurements, validating the polarization creation method. For the FUV range (96–120 nm), the polarimeter design is based on a fully reflective system, including a K-mirror modulator and a multilayer analyzer composed of B4C and MgF2. The analyzer's polarization properties have been measured at 120 nm, confirming its performance. Next steps include integrating the K-mirror and analyzer in a vacuum chamber to perform polarimetric measurements, which will allow the characterization of the FUV polarimeter's precision and validation of its design. These developments are critical for advancing the technological readiness of Pollux.

The development of UV coating technologies for Pollux and HWO focuses on optimizing materials and deposition methods for the FUV (90–120 nm), MUV (120–200 nm), and NUV (>200 nm) ranges. For the FUV channel, materials like LiF, Al, SiC, and B4C are used, with LiF being the only transparent material down to 101 nm. Broadband mirrors in the FUV are achieved using Al protected with MgF2 or AlF3, while bandpass technologies, such as multilayer coatings, are employed to enhance performance and tunability. In the MUV range, fluorides like MgF2, AlF3, and LiF are used for transparent coatings, and Al is used for reflective coatings. For the NUV and visible ranges, oxides such as Al2O3, HfO2, TiO2, and SiO2 are utilized. Advanced deposition methods, including thermal evaporation and sputtering, are employed to ensure high purity and low contamination, with in-situ measurements to validate performance [11]. These efforts aim to achieve high reflectance and spectral line discrimination, critical for Pollux's scientific objectives.

Paris Observatory and CNES are funding a development study with Safran Reosc in France for the Pollux UV to NIR dichroics. The first dichroic reflects the 120-236 nm MUV spectral range (see Figure 2) and let the longer wavelengths pass through till the second dichroic that reflects the NUV range. Initial results for the studies are expected in late 2026.

The development of diffraction gratings for Pollux involves collaboration between F. Grisé from Penn State University, the optics department from Laboratoire d'Astrophysique de Marseille (LAM) and Horiba France, focusing on optimizing echelle gratings for the FUV, MUV, and NUV channels.

The challenges for the grating's development are the high efficiency for orders between 30 and 60, the extremely low scatter and low polarization required. Grisé et al. [12] reports on the LUVOIR UV grating development and follow-up characterization and numerical performance simulations in 2025-2026 at Penn State University and LAM.Based on optical design specification, Alliguié et al. [13] have used PCGrate to simulate the grating performance during her internship at LAM. Among the study results of this internship is the optimization of MgF2 coated aluminum echelle grating for Pollux's MUV and NUV channels. Simulations showed that a 15 nm $MgF_2$ coating for MUV and 30 nm for NUV achieves an average efficiency of 70% across the spectral range, while limiting polarization variations to less than 10%. However, a sharp drop in efficiency below 120 nm was observed with $MgF_2$, confirming the need to explore alternatives such as LiF, which offers better performance in the FUV (up to 60% at 100–120 nm).

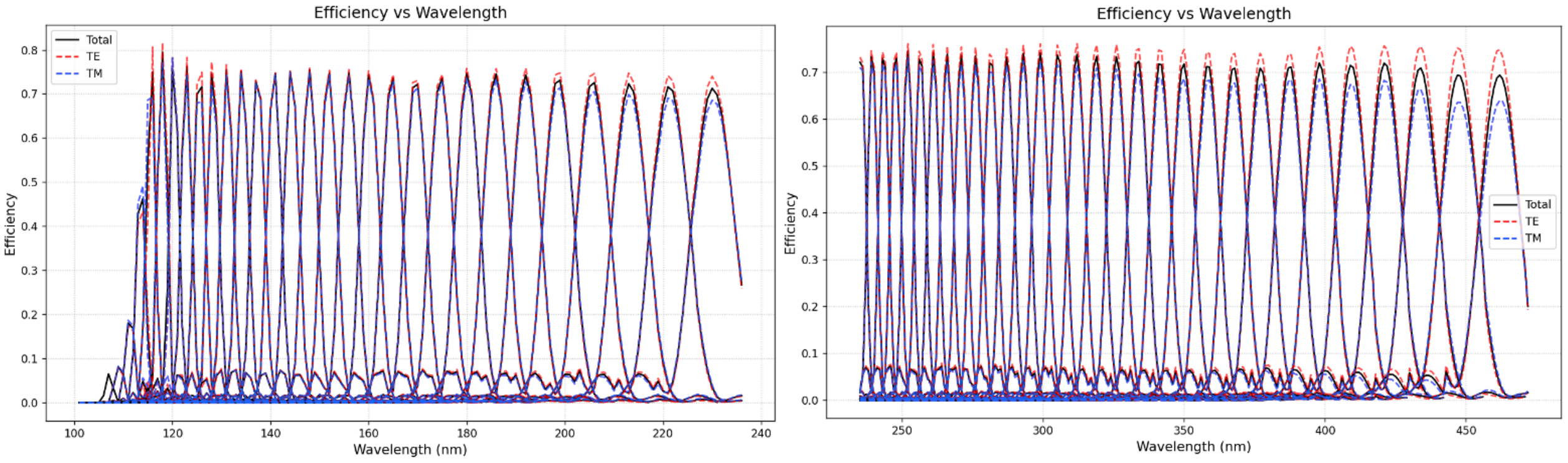

Figure 4. MUV (Left) and NUV (Right) grating efficiency simulation with optimized MgF2 coating thickness

## 8. FUTURE WORK

Pollux is a high-resolution spectropolarimeter for HWO, covering 100–1750 nm with five channels. Its design achieves R ≥ 100,000 in UV and R ≥ 65,000 in OPT/NIR, with polarimetric precision of 10^-4 to 10^-6. The international team is working on developing the science cases, building a science community, exploring trade space parameters for the architecture and the design requirements as well as producing instrument design options that are fully compatible with the HWO analytic studies EAC-3 EAC-4 and EAC-5. The instrument targets TRL 5 by 2028 and international partners are focusing on maturing technology for detectors, dichroics, coatings, and gratings. The consortium is actively working to secure funding through national agencies and the expected ESA call for HWO instrument contributions. Collaborations with US institutions for science and technological studies [12], are also being explored to strengthen the project's international partnerships. These efforts aim to ensure Pollux is well-positioned to meet the scientific and technical challenges of the HWO mission.

## ACKNOWLEDGEMENTS

The Pollux instrument studies are supported in France by the Centre National d'Etudes Spatiales (CNES).